\documentclass[amsmath,amssymb,twocolumn,nofootinbib]{revtex4-1}
\begin{document}
\title{Bianchi-I cosmological model  and crossing singularities}
\author{Alexander~Yu.~Kamenshchik}
\email{kamenshchik@bo.infn.it}
\affiliation{Dipartimento di Fisica e Astronomia, Universit\`a di Bologna\\ and INFN,  Via Irnerio 46, 40126 Bologna,
Italy,\\
L.D. Landau Institute for Theoretical Physics of the Russian
Academy of Sciences,\\
 Kosygin str. 2, 119334 Moscow, Russia}
\author{Ekaterina~O.~Pozdeeva}
\email{pozdeeva@www-hep.sinp.msu.ru}
\affiliation{Skobeltsyn Institute of Nuclear Physics, Lomonosov Moscow State University,\\
 Leninskie Gory 1, 119991, Moscow, Russia}
 \author{Alessandro~Tronconi}
\email{tronconi@bo.infn.it}
\affiliation{Dipartimento di Fisica e Astronomia, Universit\`a di Bologna\\ and INFN, Via Irnerio 46, 40126 Bologna,
Italy}
\author{Giovanni~Venturi}
\email{giovanni.venturi@bo.infn.it}
\affiliation{Dipartimento di Fisica e Astronomia, Universit\`a di Bologna\\ and INFN,  Via Irnerio 46, 40126 Bologna,
Italy}
\author{Sergey~Yu.~Vernov}
\email{svernov@theory.sinp.msu.ru}
\affiliation{Skobeltsyn Institute of Nuclear Physics, Lomonosov Moscow State University,\\
 Leninskie Gory 1, 119991, Moscow, Russia}

\begin{abstract}
We consider a rather  simple method for the description of the Big Bang -- Big Crunch cosmological singularity crossing. For the flat Friedmann universe this method gives the same results as more complicated methods, using Weyl symmetry  or the transitions between the Jordan and Einstein frames.  It is then easily generalized for the case of a Bianchi-I anisotropic universe. We  also present  early-time and late-time asymptotic solutions for a Bianchi-I universe, filled with a conformally coupled  massless scalar field.
\end{abstract}
%\pacs{98.80.Jk, 98.80.Cq, 04.20.-q, 04.20.Jb}
\maketitle

\section{Introduction}

The problem of the existence of the cosmological singularity at the beginning of  cosmological evolution has attracted the  attention of  scientists studying general relativity for a long time~\cite{Khal-Lif}.   In papers~\cite{Pen-Hawk}
the indefinite continuation of geodesics under certain conditions was shown to be impossible.
This was interpreted as pointing to the existence of a singularity in the general
solution to the Einstein equations.
The analytical behaviour of the general solutions to the Einstein equations in the
neighbourhood of a singularity was investigated in papers~\cite{BKL}.
These papers revealed the enigmatic  phenomenon of an oscillatory approach to the singularity
which has also become known as the {\it Mixmaster Universe}~\cite{Misner}.

Another type of cosmological singularity, arising for a  finite value of the cosmological scale factor, was considered in~\cite{Barrow}.
Recently, the so-called ``soft'' singularities arising for large values of the scale factor were extensively studied~\cite{soft} and the
situations for which such singularities can be crossed were found~\cite{cross,cross-we}. At the same time the idea that the Big Bang -- Big Crunch singularity can be crossed appeared very counterintuitive.

Nonetheless,  the procedure for the crossing of the Big Bang - Crunch singularity, based on the use of  Weyl symmetry, was elaborated \cite{Bars1,Bars2,Bars3,Bars4,Bars5,Bars6,Bars7}.  Using a Weyl-invariant theory, where two scalar fields conformally coupled
with gravity  were presented, the authors obtained the geodesic completeness of the corresponding spacetime. The consequence  of this geodesic completeness is the crossing of the Big Bang singularity and the emergence of  antigravity regions when using the Einstein frame.
In particular, in papers \cite{Bars3,Bars5,Bars7} the
crossing of  anisotropic singularities was considered. The authors  used the
general expressions for   Bianchi-I, Bianchi-VIII and Bianchi-IX universes. In the
vicinity of the singularity all these universes have Kasner-like behavior. This allows one to explicitly
write down  the solutions for a universe crossing such a singularity by using some finite Weyl-invariant quantities. The explicit set of geodesics for the  singularity crossing of massive and massless particles  was constructed as well.

The use of Weyl symmetry to describe the passage through the Big Crunch -- Big Bang singularity accompanied by a change of sign for the effective Newton's constant, has led to some discussion. In~\cite{polem} it was noticed that for such a passage through the singularity some curvature invariants become infinite. In paper \cite{Bars7} a counter-argument was put forward. If one has enough conditions so as to match the nonsingular quantities before and after crossing the  singularities, then the singularities can be traversed. One can say that such a treatment of the problem of  singularities crossing is in some sense common to all the attempts to describe this process.

A somewhat different approach to the problem of a cosmological singularity was developed in a series of papers~\cite{Wetterich1,Wetterich2,Wetterich3,Wetterich4,Wetterich5}. In these works the author considers the so-called variable gravity together  with the transitions between different frames and the reparametrization of the scalar field. In particular, he introduces a so-called freeze frame where the universe is very cold and slowly evolving. In the freeze picture the masses of elementary particles and the gravitational constant decrease with cosmic time, while  Newtonian attraction remains unchanged. The cosmological solution can be extrapolated to the infinite past in physical time --- the universe has no beginning. In the equivalent; however, singular,  Einstein frame cosmic history one finds the familiar Big Bang description.
Generally, the papers \cite{Wetterich1,Wetterich2,Wetterich3,Wetterich4,Wetterich5} give the impression that there is no physical singularity in Nature, but only a singularity in the choice of field coordinates. Let us note that in this scheme there is only one scalar field (just as in our present paper); however, the Hilbert-Einstein term in the action is absent, and the coupling coefficient between the scalar field squared and the scalar curvature is such  as to provide the positive sign of the effective Newtonian constant.
The instant of time which corresponds to the Big Bang in the Einstein frame (in other words ``in the Big Bang picture''),  also corresponds to the beginning of the cosmological evolution in the freeze frame, but the geometry of the spacetime is not singular: It is the field parametrization which is singular. According to the definition of the author of~\cite{Wetterich1,Wetterich2,Wetterich3,Wetterich4,Wetterich5}, the cosmology is
non-singular, if there is a frame, where the geometry is non-singular. Thus, there is an analogy between the horizon that arises due to a certain choice of the spacetime coordinates and the singularity that arises because of some choice of the field parametrization. Furthermore, the evolution occupies an infinite physical time. It is stressed that the physical time does not necessarily coincide with the proper (cosmic) time. Instead, the author
of \cite{Wetterich1,Wetterich2,Wetterich3,Wetterich4,Wetterich5} defines the physical time by counting the number of zeros of a component of the wave function. Let us note that the idea that the cosmic time is not always the best parameter for the description of  physical reality also arises in the context of the oscillatory approach to the cosmological singularity \cite{BKL}. Indeed, in this situation it is better to use the logarithmic time, which is more suitable for the description of an infinite amount of Kasner epochs and eras (see e.g. \cite{Kam-UFN} and the references therein).
The opportunity of the crossing of singularities in the Big Bang picture is not excluded in the approach \cite{Wetterich1,Wetterich2,Wetterich3,Wetterich4,Wetterich5}, but it does not arise for the concrete form of the  models considered.

 The description of the evolution of the universe through the Big Bang singularity, based on the so-called relational approach \cite{relat}, was suggested in paper \cite{relat-sing}. The authors of \cite{relat-sing} have investigated the singularities of homogenous cosmologies from the point of view of relational degrees of freedom of the gravitational field. These degrees do not depend on absolute units of length and duration --- thus they do not include the volume and extrinsic curvature. It was found that the dynamical system depending only on such variables is well posed at all physical times, even at the point that would be described as the Big Bang singularity.

A rather  radical approach to the problem of cosmological singularity, based on the Synthetic Differential Geometry
\cite{synt} was developed in paper \cite{synt-sing}. In this geometry, as the volume of the universe shrinks the curvature increases, but when the volume becomes infinitesimal, the curvature also becomes infinitesimal and the singularity is avoided.

A more traditional approach to the treatment of the black hole singularities was suggested in papers~\cite{Prester}, where
the field redefinitions were used to obtain the regular solutions.

In our preceding paper \cite{KPTVV-sing}  we   proposed a  version of the description of the crossing of singularities  in  universes filled with scalar fields. This version was based on the transitions between the Jordan and the Einstein frames.  Conformal symmetry also was used in our approach because we  implemented a particular choice of the coupling between the scalar field and the scalar curvature --- the conformal coupling. We  used a conformal coupling because in this case the relations between the parameterizations of the scalar field in the Jordan and in the Einstein frame have a simple explicit form. For simplicity we   only considered  an isotropic cosmological singularity, present in a flat Friedmann universe. We also  essentially used the relations between exact solutions of the cosmological (Friedmann and Klein-Gordon) equations in two different frames, which were studied in detail in our papers \cite{KPTVV2013,KPTVV2015}.  The main idea of  the paper~\cite{KPTVV-sing} was the following: when in the Einstein frame the universe arrives at the Big Bang -- Big Crunch singularity, from the point of view of the evolution of its counterpart in the Jordan frame its geometry is regular, but the effective Planck mass has a zero value. The solution to the equations of motion in the Jordan frame is smooth at this point and, by using the relations between the solutions of the cosmological equations in two frames, one can describe the crossing of the cosmological singularity in a uniquely determined way. The contraction is replaced by the expansion (or vice versa),
and the universe enters into the antigravity regime. Analogously, when the geometry is singular in the Jordan frame, it is regular in the Einstein frame, and using this regularity, we can describe in a well-determined way the crossing of the singularity in the Jordan frame.

It is important to note that the possibility of a change of sign of the effective gravitational  constant in the model with the scalar field conformally coupled with the scalar curvature was studied in paper~\cite{Starobinsky1981}, where the earlier suggestion, made in paper~\cite{Linde1979}, was analyzed in some detail. In paper \cite{Starobinsky1981} it was pointed out  that in a homogeneous and isotropic universe one can indeed  cross the point where the effective gravitational constant changes sign.
However, the presence of anisotropies or inhomogeneities changes the situation drastically because, when the value of the effective gravitational constant tends to zero, these anisotropies and inhomogeneities grow indefinitely. In the present paper we investigate this phenomenon, using the formalism developed in the preceding papers~\cite{KPTVV-sing,KPTVV2013,KPTVV2015}. We see why this formalism, applied naively does not work and what alternative way of crossing the singularity can be invented. We  consider a very simple cosmological model --- a Bianchi-I universe filled with a massless scalar field with vanishing potential --- because all the principal aspects of the problem under investigation can already be seen  in this case.

The structure of the paper is as follows: in the second section we discuss the relations between  models with minimally coupled and non-minimally coupled scalar fields
and show why the mechanism of the description of the crossing of the singularity, suggested in paper~\cite{KPTVV-sing}, does not work in the anisotropic Bianchi-I model.
In the third section we compare the exact Kasner solution for the case of the Bianchi-I  universe filled with a minimally coupled massless scalar field with  asymptotic solutions for the Bianchi-I universe filled with a conformally coupled massless scalar field.
In the fourth section we suggest an alternative mechanism for the description of the singularity crossing and the last section contains some brief concluding remarks.
%%%%%%%%%%%%%%%%%%%%%%%%%%

\section{Relations between  models with minimally coupled and non-minimally coupled scalar fields and the transitions between the Jordan frame and the Einstein frame}
\label{Sec2}
In this section we repeat the presentation of the formulas relating different quantities in  models with minimally and non-minimally coupled scalar fields, described in detail in papers~\cite{KPTVV2013,KPTVV2015,KPTVV-sing}. We adapt them for the case of the Bianchi-I universe.
Let us consider a cosmological model with the following action
\begin{equation}
S =\int d^4x\sqrt{-g}\left[U(\sigma)R - \frac12g^{\mu\nu}\sigma_{,\mu}\sigma_{,\nu}+V(\sigma)\right],
\label{action0}
\end{equation}
where $U(\sigma)$ and $V(\sigma)$ are differentiable  functions of the scalar field $\sigma$.

Let us make a conformal transformation of the metric,
\begin{equation}
g_{\mu\nu} = \frac{U_1}{U}\tilde{g}_{\mu\nu},
\label{conf}
\end{equation}
where $U_1$ is a constant.
We also introduce a new scalar field $\phi$ such that
\begin{equation}
\frac{d\phi}{d\sigma} = \frac{\sqrt{U_1(U+3U'^2)}}{U}
\quad\Rightarrow\quad
\phi =\! \int\! \frac{\sqrt{U_1(U+3U'^2)}}{U} d\sigma.
\label{scal1}
\end{equation}

The action (\ref{action0}) then becomes the action for a minimally coupled scalar field:
\begin{equation}
S =\int d^4x\sqrt{-\tilde{g}}\left[U_1R(\tilde{g}) - \frac12\tilde{g}^{\mu\nu}\phi_{,\mu}\phi_{,\nu}+W(\phi)\right] ,
\label{action1e}
\end{equation}
where
\begin{equation}
W(\phi) = \frac{U_1^2 V(\sigma(\phi))}{U^2(\sigma(\phi))}.
\label{poten}
\end{equation}
The transformation from the metric $g_{\mu\nu}$ and the scalar field $\sigma$ to the metric $\tilde{g}_{\mu\nu}$ and the scalar field $\phi$, described by the action (\ref{action1}), is  the transformation from the Jordan frame to the Einstein frame.

Let us emphasise that the formulae (\ref{conf})--(\ref{poten}) are valid for any metric.
Now we  consider a Bianchi-I universe with the interval
\begin{equation}
\begin{split}
ds^2&=\tilde{N}^2(\tau)d\tau^2\\&{}-\tilde{a}^2(\tau)(e^{2\beta_1(\tau)}dx_1^2+e^{2\beta_2(\tau)}dx_2^2+e^{2\beta_3(\tau)}dx_3^2).
\end{split}
\label{Bianchi-I}
\end{equation}
Here $\tilde{N}$ is the lapse function, $\tilde{a}$ is the scale factor, and $\beta_i$ are anisotropy factors, such that
\begin{equation}
\beta_1+\beta_2+\beta_3=0.
\label{anisot}
\end{equation}

The conformal transformation (\ref{conf}) preserves the Bianchi-I form of the metric (\ref{Bianchi-I})  with the following changes for the functions
\begin{equation}
\tilde{N} = \sqrt{\frac{U}{U_1}}N,\qquad \tilde{a} =  \sqrt{\frac{U}{U_1}}a.
\label{Na}
\end{equation}
The anisotropy factors $\beta_i$ are unchanged.

There is  also another useful relation, connecting the cosmic time parameters in different frames~\cite{KPTVV-sing}:
\begin{equation}
t =  \int \sqrt{\frac{U_1}{U}}d\tilde{t}.
\label{time-frame}
\end{equation}

On varying the action (\ref{action0}) and substituting the Bianchi-I metric (\ref{Bianchi-I}) it is easy to get the corresponding Einstein equations (see, for example, Ref.~\cite{Kamenshchik:2016atu}).

For the purposes of this paper it is more convenient to use the minisuperspace action for the metric (\ref{Bianchi-I}) and the spatially homogeneous
scalar field~$\phi$.

Taking into account the fact that the anisotropy  coefficients $\beta_i$ are not independent [Eq.~(\ref{anisot})], it is convenient to introduce two independent coefficients $\alpha_1$ and $\alpha_2$ such that
\begin{equation}
\label{anisot2}
\begin{split}
\beta_1 &= \frac{1}{\sqrt{6}}\alpha_1+\frac{1}{\sqrt{2}}\alpha_2, \\
\beta_2 &= \frac{1}{\sqrt{6}}\alpha_1-\frac{1}{\sqrt{2}}\alpha_2, \\
\beta_3 &={}-\frac{2}{\sqrt{6}}\alpha_1.
\end{split}
\end{equation}

A direct calculation gives the Ricci scalar
\begin{equation}
R=\frac{1}{\tilde{N}^2}\left[6\frac{\dot{\tilde{N}}\dot{\tilde{a}}}{\tilde{N}\tilde{a}}
-6\frac{\ddot{\tilde{a}}}{\tilde{a}}-6\frac{\dot{\tilde{a}}^2}{\tilde{a}^2}
-\dot{\alpha}^2_1-\dot{\alpha}^2_2\right].
\label{curvature}
\end{equation}

The Einstein frame action (\ref{action1e}) generates the following minisuperspace action
\begin{equation}
\label{action}
\begin{split}
S&=\!\int\!  d\tau d^3x\left[6\frac{U_1\tilde{a}^3}{\tilde{N}}\left(\frac{\dot{\tilde{N}}\dot{\tilde{a}}}{\tilde{N}\tilde{a}}
-\frac{\ddot{\tilde{a}}}{\tilde{a}}-\frac{\dot{\tilde{a}}^2}{\tilde{a}^2}\right)\right. \\
&\left.{}-\frac{U_1\tilde{a}^3}{\tilde{N}}\left(\dot{\alpha}^2_1+\dot{\alpha}^2_2\right)
-\frac{\tilde{a}^3\dot{\phi}^2}{2\tilde{N}}+\tilde{a}^3\tilde{N}W(\phi)\right]\!.
\end{split}
\end{equation}

The main goal of this section is to check whether the method proposed for the isotropic case in~\cite{KPTVV-sing} can be applied in the case of the Bianchi-I metric. To do this we  consider the case of a vanishing potential $W=0$.
Integrating by parts we arrive at
\begin{equation}
S=\!\int\! d\tau d^3x \frac{1}{\tilde{N}}\left[6U_1\dot{\tilde{a}}^2\tilde{a}-U_1\tilde{a}^3\left(\dot{\alpha}^2_1+\dot{\alpha}^2_2\right)
-\frac{\tilde{a}^3\dot{\phi}^2}{2}\right].
\label{action1}
\end{equation}
Varying the action (\ref{action1}) with respect to the variables $\phi$ and $\alpha_i$, it is easy to obtain
\begin{equation}
\frac{d}{d\tau}\left(\frac{\tilde{a}^3}{\tilde{N}}\dot{\alpha_i}\right)=0,\quad\Rightarrow\quad \dot{\alpha_i}=\frac{\tilde{N}\alpha_{i0}}{\tilde{a}^3},\ i=1,2,
\label{beta}
\end{equation}
or, in other terms,
\begin{equation}
\frac{d}{d\tau}\left(\frac{\tilde{a}^3}{\tilde{N}}\dot{\beta_i}\right)=0,\quad\Rightarrow\quad \dot{\beta_i}=\frac{\tilde{N}\beta_{i0}}{\tilde{a}^3},\ i=1,2,3
\label{beta0}
\end{equation}
and
\begin{equation}
\frac{d}{d\tau}\left(\frac{\tilde{a}^3}{\tilde{N}}\dot{\phi}\right)=0,\quad\Rightarrow\quad\dot{\phi}=\frac{\tilde{N}\psi_1}{\tilde{a}^3},
\label{phi}
\end{equation}
where $\alpha_{i0}$, $\beta_{i0}$ and $\psi_1$ are constants. The connections between  $\alpha_{i0}$ and $\beta_{i0}$ are given by~(\ref{anisot2}).

The variation of the action (\ref{action1}) with respect to the lapse function $\tilde{N}$ gives an analog of the Friedmann equation
\begin{equation}
6U_1\frac{\dot{\tilde{a}}^2}{\tilde{a}^2}=U_1\left(\dot{\alpha}_1^2+\dot{\alpha}_2^2\right)+\frac{\dot{\phi}^2}{2}.
\label{Fried}
\end{equation}
 Choosing the lapse function $\tilde{N}=1$ we identify the parametric time $\tau$ with the cosmic time $\tilde{t}$ in the Einstein frame, and using Eqs.~(\ref{anisot2}),
 (\ref{beta}), and (\ref{phi}),  we obtain
\begin{equation}
6U_1\frac{\dot{\tilde{a}}^2}{\tilde{a}^2}=\frac{2U_1\theta_0+\psi_1^2}{2\tilde{a}^6},
\label{Fried1}
\end{equation}
 where
 \begin{equation}
 \theta_0=\beta_{10}^2+\beta_{20}^2+\beta_{30}^2=\alpha_{10}^2+\alpha_{20}^2.
 \label{theta}
 \end{equation}

 Equation~(\ref{Fried1}) can be easily integrated and we have
 \begin{equation}
 \tilde{a}(\tilde{t})= \left(\frac{3\theta_0}{2}+\frac{3\psi_1^2}{4U_1}\right)^{\frac{1}{6}}(\tilde{t}-\tilde{t}_0)^{1/3}.
 \label{Fried2}
 \end{equation}
 Obviously,  there is a Big Bang -- Big Crunch singularity at the moment $\tilde{t}=0$ if we put the integration constant $\tilde{t}_0=0$.
Substituting~(\ref{Fried2}) with $\tilde{t}_0=0$ into Eq.~(\ref{phi}), we obtain
 \begin{equation}
 \phi=\frac{\psi_1}{\left(\frac{3\theta_0}{2}+\frac{3\psi_1^2}{4U_1}\right)^{\frac12}}\ln (\tilde{t}) +\phi_0,
 \label{phi1}
 \end{equation}
 where  $\phi_0$ is a  constant.
 Let us note that for the isotropic case when $\theta_0=0$, considered in paper~\cite{KPTVV-sing},
 \begin{equation}
 \phi=\pm\sqrt{\frac{4U_1}{3}}\ln(\tilde{t})+\phi_0,
 \label{phi2}
 \end{equation}
 i.e. the coefficient in front of the $\ln(\tilde{t})$ is not free but is rigidly defined up to a sign.

 Let us now try to apply the trick, used in the preceding paper \cite{KPTVV-sing} to find a prescription for the
 matching of the quantities, characterizing the cosmological evolution before and after the singularity. We
   considered the Jordan frame with the function $U$, which included the Hilbert-Einstein term plus the term describing the nonminimal conformal coupling between the scalar curvature and scalar field:
 \begin{equation}
 U_c=U_0-\frac{1}{12}\sigma^2,
 \label{nonmin-conf}
 \end{equation}
where $U_0$ is a positive constant.

Let us note that the models with $U_c$ have been  actively studied in  recent years~\cite{KPTVV2013,KPTVV2015,Boisseau:2015hqa,BNOS,PSTV2016}.
 Taking into account that
\begin{equation}\label{gsh}
U_c+3{U'_c}^2=U_0,
\end{equation}
 we get from~(\ref{scal1})
\begin{equation}
\label{phic}
\phi=\! \int\! \frac{\sqrt{U_1U_0}}{U_c} d\sigma .
\end{equation}
Hence, $\phi$ is real for $U_1>0$ and is imaginary for $U_1<0$. In the models we consider, the imaginary scalar field can be treated as a real scalar field with a phantom kinetic term.
At $U_1>0$ and $U_c>0$ we get
\begin{equation}
\phi = \sqrt{3U_1}\ln \left[\frac{\sqrt{12U_0}+\sigma}{\sqrt{12U_0}-\sigma}\right]
\label{connection_c0}
\end{equation}
and vice versa
\begin{equation}
\sigma = \sqrt{12U_0}\tanh\left[ \frac{\phi}{\sqrt{12U_1}}\right].
\label{connection_c}
\end{equation}

When $-\infty < \phi < \infty$, from Eq.~(\ref{connection_c}) it follows that $U_c>0$. When $|\phi| \rightarrow \infty$, we usually have a singularity in the Einstein frame and the value of the field $\sigma$ in the Jordan frame
tends to its limiting value $|\sigma| = \sqrt{12U_0}$ and, hence, $U_c\rightarrow0$.

However, from the point of view of the proper dynamics of the field $\sigma$ in the Jordan frame, there is nothing that prevents it from crossing the value $\sigma = \pm \sqrt{12U_0}$. Moreover, also in the isotropic case  the scale factor $a$ differs from zero when  $\sigma = \pm \sqrt{12U_0}$. This has allowed  us to describe the crossing of the singularity as  was done in paper \cite{KPTVV-sing}.

We now briefly show  how it worked and why it does not work in the anisotropic case.
First of all, let us note that from Eqs.~(\ref{Na}), (\ref{nonmin-conf}), and  (\ref{connection_c}) it follows that the scale factor $a$ in the Jordan frame behaves as
\begin{equation}
a =\tilde{a}\sqrt{\frac{U_1}{U_0}}\cosh\left(\frac{\phi}{\sqrt{12U_1}}\right).
\label{Jordan-scale}
\end{equation}
For the case of an isotropic universe, by substituting into Eq.~(\ref{Jordan-scale}) the expression (\ref{phi2}), we see that
when the universe tends to the singularity in the Einstein frame, i.e. when $\tilde{t} \rightarrow 0$, the scale factor in the Jordan frame behaves as
\begin{equation}
a \sim \tilde{t}^{\frac13}\left(\tilde{t}^{\frac13}+\tilde{t}^{-\frac13}\right) \rightarrow const \neq 0.
\label{Jordan-scale1}
\end{equation}
Thus, the evolution in the Jordan frame is regular and we can use this fact to describe the crossing of the Big Bang -- Big Crunch singularity in the Einstein frame, as  was done in paper \cite{KPTVV-sing}. Such a description  boils down to the fact that if one considers the expansion of the universe from the Big Bang with normal gravity driven by the standard scalar field, the continuation backward in time shows that it was preceded by the contraction towards a Big Crunch singularity in the antigravity regime, driven by a phantom scalar field with a negative kinetic term. Let us now consider  the anisotropic case, when the constant $\theta_0$ in  formula (\ref{phi1}) is different from zero. Then,  substituting the expression (\ref{phi1}) into Eq.~(\ref{Jordan-scale}), one obtains
\begin{equation}
a \sim \tilde{t}^{\frac13}\left(\tilde{t}^{\gamma}+\tilde{t}^{-\gamma}\right) \rightarrow 0,
\label{Jordan-scale2}
\end{equation}
where
\begin{equation}
\gamma=\frac{\psi_1}{3\sqrt{\psi_1^2 +2\theta_0U_1}} < \frac13.
\end{equation}

Thus, if one also encounters the Big Bang singularity  in the Jordan frame, the algorithm used in paper~\cite{KPTVV-sing} cannot be applied directly.

\section{Bianchi-I universes filled  with minimally coupled and conformally coupled massless scalar fields}

In the preceding section we have seen that the transitions between the Einstein frame and the Jordan frame do not help to describe crossing  the Big Bang -- Big Crunch singularity. Nevertheless,  comparison between the cosmological evolutions in the Einstein frame and in the Jordan frame is of interest by itself. We discuss this in the present section, while in the next section we return to the question of the singularity crossing.

Let us first consider  the exact solution for a Bianchi-I universe
filled with a minimally coupled massless scalar field.
In the preceding section we have already found the expression for the scale factor $\tilde{a}(\tilde{t})$ given by Eq. (\ref{Fried2}). Substituting this expression into Eq.~(\ref{beta0}) and integrating it with respect to the cosmic time $\tilde{t}$, we find that
\begin{equation}
\beta_i=\frac{\beta_{i0}}{\sqrt{\frac{3\theta_0}{2}+\frac{3\psi_1^2}{4U_1}}}\ln(\tilde{t})+const.
\label{beta2}
\end{equation}

Now substituting the expressions (\ref{Fried2}) and (\ref{beta2}) into Eq.~(\ref{Bianchi-I}), we can represent the exact solution for the Bianchi-I universe in a standard Kasner-like form~\cite{Khal-Lif}:
\begin{equation}
ds^2=d\tilde{t}^2-a_1^2\tilde{t}^{2p_1}dx_1^2-a_2^2\tilde{t}^{2p_2}dx_2^2-a_3^2\tilde{t}^{2p_3}dx_3^2,
\label{Kasner}
\end{equation}
where the Kasner indices are given by the expressions
\begin{equation}
p_i=\frac13+\frac{\beta_{i0}}{\sqrt{\frac{3\theta_0}{2}+\frac{3\psi_1^2}{4U_1}}}
\label{Kasner1}
\end{equation}
and satisfy the following relations:
\begin{equation}
p_1+p_2+p_3 = 1,
\label{Kasner2}
\end{equation}
\begin{equation}
p_1^2+p_2^2+p_3^2 = \frac13+\frac{\theta_0}{\frac{3\theta_0}{2}+\frac{3\psi_1^2}{4U_1}} < 1.
\label{Kasner3}
\end{equation}

Let us note that while  relation (\ref{Kasner2}) coincides with the known relation for the Kasner indices in an empty Bianchi-I universe \cite{Khal-Lif},
the relation (\ref{Kasner3}) is modified by the presence of the massless scalar field. The fact that the presence of the massless scalar field modifies the second Kasner relation is rather well known~\cite{Bel-Khal,Khal-Kam}.

Let us now  turn to the consideration of the case of a massless scalar field, conformally coupled to gravity. One can show that in this case the main equations of motion
 are~\cite{Kamenshchik:2016atu}
\begin{equation}
6\frac{\dot{a}^2}{a^2}U_c +6\frac{\dot{a}}{a}\dot{\sigma}U_c'=\frac12\dot{\sigma}^2+U_c(\dot{\bar{\beta}}_1^2+\dot{\bar{\beta}}_2^2+\dot{\bar{\beta}}_3^2),
\label{nonmin-F}
\end{equation}
\begin{equation}
\label{Bianchi_trace}
     2U_c\frac{\ddot{a}}{a}+4U_c \frac{\dot{a}^2}{a^2}+U_c''\dot{\sigma}^2+U_c'
     \left[\ddot{\sigma}+5\frac{\dot{a}}{a}\dot{\sigma}\right]=0\,,
\end{equation}
\begin{equation}
\ddot{\sigma}+3\frac{\dot{a}}{a}\dot{\sigma}-RU_c'= 0\,,
\label{KGB1}
\end{equation}
 where the dots mean the derivative with respect to the cosmic time $t$ and the primes mean the derivative with respect to the scalar field $\sigma$. The functions $\bar{\beta}_i$ are the anisotropy factors in the Jordan frame. The function $N=1$.

Straightforward calculations [see~(\ref{curvature})] show that $R=0$, and Eq.~(\ref{KGB1}) gives
\begin{equation}
\dot{\sigma}=\frac{\sigma_0}{a^3}.
\label{nonmin-s}
\end{equation}

From Eqs.~(\ref{Na}) and (\ref{beta0}) it follows that
 \begin{equation}
 \dot{\bar{\beta}}_i=\frac{\bar{\beta}_{i0}}{U_ca^3}.
 \label{nonmin-b}
\end{equation}

The system of equations (\ref{nonmin-F}), (\ref{nonmin-s}), and (\ref{nonmin-b}) is still rather complicated.
A  possible way to simplify it is to assume that
\begin{equation}
\label{F}
    \dot\sigma=F(\sigma),
\end{equation}
where $F(\sigma)$ is a differentiable function that should be found (see~\cite{KTVV2013}).
From Eq.~\eqref{KGB1} we get the Hubble parameter as a function of $\sigma$:
 \begin{equation}\label{Hsigma}
    H\equiv \frac{\dot{a}}{a}={}-\frac{\ddot\sigma}{3\dot\sigma}={}-\frac13F'(\sigma).
 \end{equation}

 Now Eq.~(\ref{Bianchi_trace}) has the following form:
 \begin{equation}
\label{Bianchi_traceF}
     2U_c\left({F'}^2-F''F\right)-\frac{1}{2}F^2+\frac{\sigma}{3}FF'=0.
\end{equation}
Equation~(\ref{Bianchi_traceF}) has the following solution:
\begin{equation}
F=C_1 U_c^{3/2}e^{C_2 {\rm arctanh}(\sqrt{3}\,\sigma/(6\sqrt{U_0}))},
\end{equation}
where $C_1$ and $C_2$ are constants.

However, this general form is not too useful for our purposes. It is more interesting to
study the asymptotic behavior of the universe in the vicinity of the Big Bang singularity. Here, some observations made at the end of the preceding section can be rather useful. First, we have seen that for $\tilde{a} \rightarrow 0$  its counterpart $a$ in the Jordan frame also tends to zero, and the universe encounters the Big Bang singularity simultaneously in both the frames, in contrast to the isotropic Friedmann case considered in detail in our preceding paper \cite{KPTVV-sing}. Furthermore, the scalar field $\phi$ tends to $\pm \infty$, and as one can see from Eqs.~(\ref{connection_c0}) and (\ref{connection_c}) the field $\sigma$ tends to $\pm \sqrt{12U_0}$. Thus, we can consider the situation for which, as $t \rightarrow 0_+$, the scalar field $\sigma$ behaves as
\begin{equation}
\sigma ={}-\sqrt{12U_0} +\sigma_1 t^{\delta},
\label{nonmin-s1}
\end{equation}
while the scale factor behaves as
\begin{equation}
a = a_0 t^{\eta},
\label{eta}
\end{equation}
where
\begin{equation}
0 < \delta < 1
\label{delta}
\end{equation}
and
\begin{equation}
\eta > 0.
\label{eta1}
\end{equation}

Substituting the expressions (\ref{nonmin-s1}) and (\ref{eta}) into Eq.~(\ref{nonmin-s}), we find that
\begin{equation}
\delta = 1- 3\eta
\label{delta-eta}
\end{equation}
and
\begin{equation}
\sigma_0=a_0^3\delta\sigma_1.
\label{sigma-a}
\end{equation}
From the relation (\ref{delta-eta}) and the inequality (\ref{delta}) it follows that
\begin{equation}
\eta < \frac13.
\label{eta2}
\end{equation}
Substituting the expressions (\ref{nonmin-b}), (\ref{nonmin-s1}), (\ref{eta}), and (\ref{delta-eta})  into Eq.~(\ref{nonmin-F}), we see that
the first and second terms on the left-hand side of this equation and the second term on its right-hand side behave as $t^{\delta-2}$ while
the first term on the right-hand side of this equation behaves as $t^{2\delta-2}$ and, hence, is subleading. Then, comparing the coefficients for the leading terms
in Eq. (\ref{nonmin-F}), we obtain the following relation:
\begin{equation}
\bar{\theta}_0\equiv \bar{\beta}_{10}^2+\bar{\beta}_{20}^2+\bar{\beta}_{30}^2=2U_0\sigma_1^2a_0^6\eta(1-2\eta).
\label{relat}
\end{equation}
Now, integrating Eq.~(\ref{nonmin-b}) and taking into account the relations obtained above, we obtain the following expression for the anisotropic factors:
\begin{equation}
\bar{\beta}_i = \frac{\bar{\beta}_{i0}\sqrt{6\eta(1-2\eta)}}{\sqrt{\bar{\theta}_0}}\ln t + const.
\label{beta-bar}
\end{equation}
This means that our Bianchi-I universe in the vicinity of the singularity  also has  the Kasner-like form of the metric with the Kasner indices
\begin{equation}
p_i=\eta+\frac{\bar{\beta}_{i0}\sqrt{6\eta(1-2\eta)}}{\sqrt{\bar{\theta}_0}}.
\label{Kasner-new}
\end{equation}
The new relations between the Kasner indices are
\begin{equation}
p_1+p_2+p_3=3\eta < 1
\label{Kasner-new1}
\end{equation}
and
\begin{equation}
p_1^2+p_2^2+p_3^2 =3\eta(2-3\eta).
\label{Kasner-new2}
\end{equation}
We see that these Kasner relations depend only on the parameter $\eta$. We can establish a relation between this parameter $\eta$ and the parameters describing the evolution of the universe in the Einstein frame. Let us return to Eq.~(\ref{Jordan-scale2}). In the vicinity of the singularity when $\tilde{t} \rightarrow 0$, we have
\begin{equation}
a \sim \tilde{t}^{\frac13-\gamma}.
\label{J-E}
\end{equation}
Using the relation (\ref{time-frame}), we find that
\begin{equation}
\tilde{t} \sim t^{\frac{1}{1-\gamma}}.
\label{J-E1}
\end{equation}
Hence,
\begin{equation}
a \sim t^{\frac{1/3-\gamma}{1-\gamma}}
\label{J-E2}
\end{equation}
or, in other words,
\begin{equation}
\eta = \frac{1-3\gamma}{3(1-\gamma)} < \frac13.
\label{eta-EJ}
\end{equation}

We can now analyse what happens at the late stage of the evolution, when $t \rightarrow \infty$. In this case the relation between $\tilde{t}$ and $t$
is
\begin{equation}
\tilde{t} \sim t^{\frac{1}{1+\gamma}},
\label{J-E1}
\end{equation}
 and when $\tilde{t}$ tends to $+\infty$, the cosmic time parameter $t$ does  as well. The scalar field $\sigma \rightarrow +\sqrt{12U_0}$, and the formula
(\ref{nonmin-s1}) is replaced by
\begin{equation}
\sigma = +\sqrt{12U_0} +\sigma_1 t^{\delta}.
\label{nonmin-s10}
\end{equation}
The relations (\ref{delta-eta}), (\ref{sigma-a}), (\ref{relat}), (\ref{beta-bar})  and (\ref{Kasner-new}) maintain  their form, with the only difference being that the exponent $\delta$ should be negative  and, hence,
\begin{equation}
\eta > \frac13.
\label{eta-late}
\end{equation}
The late-time evolution of the universe still has a Kasner form and
the identities (\ref{Kasner-new1}) and (\ref{Kasner-new2}), relating the Kasner indices, are also valid.
The expression for the exponent $\eta$ [Eq.~(\ref{eta-EJ})] is substituted by
\begin{equation}
\eta = \frac{1+3{\gamma}}{3(1+\gamma)} > \frac13.
\label{eta-EJ10}
\end{equation}
The values of the Kasner indices can be represented by the points lying on the intersection of the plane (\ref{Kasner-new1}) and the sphere (\ref{Kasner-new2}).
The opportunity of having all three Kasner indices positive depends on the ratio
\begin{equation}
\frac{(p_1+p_2+p_2)^2}{p_1^2+p_2^2+p_3^2}.
\label{ratio}
\end{equation}
In paper \cite{KTVV} it was shown that the combination of three positive Kasner indices is possible if this ratio is greater than $1$.
Thus, substituting into the ratio (\ref{ratio}) the expressions (\ref{Kasner-new1}) and (\ref{Kasner-new2}) and solving the inequality
\begin{equation}
\frac{(p_1+p_2+p_2)^2}{p_1^2+p_2^2+p_3^2} =\frac{9\eta^2}{3\eta(2-3\eta)} > 1,
\label{ratio1}
\end{equation}
we find
\begin{equation*}
\eta > \frac13.
\end{equation*}

We see that this situation  can  only be realized for a late-time asymptotic regime.
If we wish to have all three Kasner indices necessarily positive for a given value of the ratio, such a ratio  should be greater than $2$~\cite{KTVV}.
Thus, solving the inequality
\begin{equation}
\frac{(p_1+p_2+p_2)^2}{p_1^2+p_2^2+p_3^2} =\frac{9\eta^2}{3\eta(2-3\eta)} > 2,
\label{ratio2}
\end{equation}
we obtain
\begin{equation*}
\eta> \frac49.
\end{equation*}

 Generally, we can say that the higher the value of the index $\eta$, the more isotropic the solution.
Thus, the late-time Kasner-type universe is more isotropic than the early time (close to the Big Bang singularity) Kasner-type universe. Let us note, however, that this kind of  isotropization is a rather weak one. For example, if one considers a Bianchi-I universe filled with dust-like matter, then at the beginning of the evolution we have a standard Kasner regime, while at a late time, we have  an isotropic Friedmann flat universe~\cite{Heck-Schuck,Khal-Kam}. Nothing similar happens in the model, which is considered in the present section.

\section{Mixing between geometrical and matter degrees of freedom and the singularity crossing}

The descriptions of the singularity crossing done in papers~\cite{Bars1,Bars2,Bars3,Bars4,Bars5,Bars6,Bars7} and in our preceding paper
\cite{KPTVV-sing} essentially used the mixing between geometrical degrees of freedom and the matter degrees of freedom associated  with the presence in the models under consideration of  scalar fields. In both approaches conformal symmetry was also used. In the preceding section we showed that the trick implemented in paper \cite{KPTVV-sing}
for the description of the crossing of the Big Bang -- Big Crunch singularity in the isotropic Friedmann universe does not work for an anisotropic Bianchi-I universe. Here we show that a simpler way of mixing the gravitational and scalar field degrees of freedom allows one not only to describe the crossing of the singularity in the Friedmann universe, but it is also rather easily generalizable for the case of an anisotropic universe.

Let us first consider the case without anisotropy. In this case the action (\ref{action1}) can be written in the following form:
\begin{equation}
S=\int d\tau d^3x \frac{1}{\tilde{N}}\left(6U_1\dot{\tilde{a}}^2\tilde{a}-\frac{\tilde{a}^3\dot{\phi}^2}{2}\right).
\label{action1f}
\end{equation}

It will be convenient now to introduce the functions
\begin{equation}
r \equiv \frac{4\sqrt{U_1}}{\sqrt{3}}\tilde{a}^{\frac32}
\label{r}
\end{equation}
and
\begin{equation}
\varphi \equiv \frac{\sqrt{3}}{4\sqrt{U_1}}\phi.
\label{varphi}
\end{equation}
We choose the lapse function $\tilde{N} =1$ and cosmic time parameter $\tilde{t}$.
The Lagrangian of our action in these variables is simply
\begin{equation}
L = \frac12\dot{r}^2-\frac12r^2\dot{\varphi}^2.
\label{Lagrange}
\end{equation}
Introducing the Cartesian variables
\begin{equation}
x=r\cosh\varphi
\label{x}
\end{equation}
and
\begin{equation}
y=r\sinh\varphi,
\label{y}
\end{equation}
we can rewrite the Lagrangian~(\ref{Lagrange}) as
\begin{equation}
L = \frac12\dot{x}^2-\frac12\dot{y}^2.
\label{Lagrange1}
\end{equation}
It is easy to see that in terms of these new variables $x$ and $y$, the constraint equation~(\ref{Fried})
is
\begin{equation}
\dot{x}^2-\dot{y}^2=0.
\label{constraint}
\end{equation}
Let us note that similar changes of variables were used largely for the construction of exact solutions in the cosmological models, including scalar fields~\cite{change}.
Remarkably, the relations (\ref{x}) and (\ref{y}) are valid only for $x > 0,\ x > |y|$. However, the dynamics of the Cartesian coordinates $x$ and $y$, implied by the Lagrangian
(\ref{Lagrange1}), brings them outside  this region  of the plane $(x,y)$. We use this fact for the description of the singularity crossing. It is rather similar to the procedure of the geodesic completion of the coordinates for the Schwarzschild solution, obtained by using the Kruskal coordinates on the plane~\cite{Kruskal}.

The inverse relations connecting the scale factor and the scalar field with the variables $x$ and $y$ are
\begin{equation}
\tilde{a}^3 = \frac{3(x^2-y^2)}{16U_1}
\label{a-xy}
\end{equation}
and
\begin{equation}
\phi=\frac{4\sqrt{U_1}}{\sqrt{3}}{\rm arctanh}\left(\frac{x}{y}\right).
\label{phi-xy}
\end{equation}
The equations of motion for the variables $x$ and $y$ are
\begin{equation*}\ddot{x} = 0,\quad  \ddot{y} = 0.
\end{equation*}

Their solutions are, obviously,
\begin{equation}
x = x_1\tilde{t}+x_0,
\label{x-sol}
\end{equation}
\begin{equation}
y = y_1\tilde{t}+y_0.
\label{y-sol}
\end{equation}
These solutions are quite regular for any value of the time parameter $\tilde{t}$.
The constraint (\ref{constraint}) implies that
\begin{equation}
x_1^2=y_1^2.
\label{constraint1}
\end{equation}
Let us now calculate  $\tilde{a}^3$:
\begin{equation}
\tilde{a}^3 = \frac{3}{16U_1}\left((x_1^2-y_1^2)\tilde{t}^2+2(x_1x_0-y_1y_0)\tilde{t}+(x_0^2-y_0^2)\right).
\label{a-xy1}
\end{equation}
Because of the constraint (\ref{constraint1}), the quadratic in time term in (\ref{a-xy}) disappears. Then, without loss of generality, we can choose the constants as follows:
\begin{equation}
x_0=y_0=A>0, \ x_1=-y_1=B > 0.
\label{a-xy2}
\end{equation}
Then,
\begin{equation}
\tilde{a}^3=\frac{3AB\tilde{t}}{4U_1}.
\label{a-xy3}
\end{equation}
For $\tilde{t} > 0$ the function $\tilde{a}$ is positive, and we have the Big Bang singularity as $\tilde{t} \rightarrow 0_+$. How can we interpret the meaning of the expression (\ref{a-xy3}) for $\tilde{t} < 0$? We see that the only way out is to  switch  the sign of  $U_1$ in the denominator of the expression (\ref{a-xy3}). This means that crossing the singularity leads to the antigravity regime.
One can  also see that when $\tilde{t}$ becomes negative and $|x| < |y|$, the variable $\varphi$ acquires an imaginary shift $\frac{i\pi}{2}$. This does not change the behavior of the time derivative of $\varphi$. However, as one can see from Eq.~(\ref{phi-xy}), the relation between an initial scalar field $\phi$ and the variable $\varphi$ includes  $\sqrt{U_1}$,
which  becomes imaginary when $U_1$ switches sign. This can be interpreted as a transformation of the standard scalar field into a phantom scalar field with a negative sign of the kinetic term. Thus, the  simple way of describing   the crossing of the singularity based on the parametrization of the metric and scalar field variables, introduced in this section,
gives the same result as a description based on the transformation between Einstein and Jordan frames, discussed in detail in paper~\cite{KPTVV-sing} and in the preceding section of the present paper.

How can we generalize the above considerations to the case when the anisotropy term is present?
First of all let us note that the Lagrangian for such a case can be written as a Lagrangian of a model, describing a Friedmann universe, filled with a triplet of scalar fields:
\begin{equation}
L = \frac12\dot{r}^2-\frac12r^2\left(\dot{\varphi}^2 +\dot{\varphi}_1^2+\dot{\varphi}_2^2\right).
\label{Lagrange3}
\end{equation}
Here the scalar fields $\varphi_1$ and $\varphi_2$ are
\begin{equation}
\varphi_1 = \sqrt{\frac38}\alpha_1,\ \varphi_2 = \sqrt{\frac38}\alpha_2.
\label{varphi-anis}
\end{equation}
Now, we can again  consider the plane $(x,y)$ such that
\begin{equation}
x = r\cosh\Phi,
\label{x-new}
\end{equation}
\begin{equation}
y=r\sinh\Phi,
\label{y-new}
\end{equation}
where a new hyperbolic angle $\Phi$ is defined as
\begin{equation}
\Phi \equiv \int d\tilde{t}\sqrt{\dot{\varphi}_1^2+\dot{\varphi}_2^2+\dot{\varphi}^2}.
\label{Phi}
\end{equation}
We have managed to reduce a four-dimensional problem to the old two-dimensional problem, using the fact that the variables $\alpha_1, \alpha_2$ and
$\phi$ enter into the equation of motion for the scale factor $a$ only through the squares of their time derivatives. Thus, the problem is reduced to that of the isotropic case and the transition to the part of the plane, where $|x| < |y|$ at $\tilde{t} < 0$, gives us an anisotropic universe contracting towards the Big Crunch singularity in the antigravity regime.
It is driven by the phantom scalar field.  Thus, we see that the behavior of the scale factor $a$ before and after  crossing the cosmological singularity can be matched by using the transition to the new coordinates $x$ and $y$, which  mix  the geometrical and scalar field variables in a particular way. What is about the anisotropic factors
$\beta_i$ (or $\alpha_i$, if we would like to consider only independent variables)? In order to describe their behavior, it is enough to fix the constants $\beta_{i0}$ and to use the formulas (\ref{beta0}), relating the anisotropy factors and the scale factor. Thus, the method of the description of the singularity crossing, suggested in this section, also works  for an anisotropic universe.

\section{Concluding remarks}

In this paper we have suggested a rather  simple method of describing  the Big Bang -- Big Crunch cosmological singularity crossing. For the flat Friedmann universe this method gives the same results as  more complicated methods, using the Weyl symmetry as in~\cite{Bars1,Bars2,Bars3,Bars4,Bars5,Bars6,Bars7} or the transitions between the Jordan and Einstein frames as  was done in our preceding paper \cite{KPTVV-sing}. It is then easily generalized to the case of a Bianchi-I anisotropic universe.
One can say  that there is a certain analogy between the appearance of  horizons as a consequence of a certain choice of the spacetime coordinates and the appearance of the singularity as a consequence  of the choice of the field parametrizations. Even if the appearance of the horizon in the Schwarzschild metric is a result of a standard choice of the coordinates, its consequences for a distant observer are quite real: He does not receive  signals coming from the region which lies behind the horizon. The introduction of  Kruskal coordinates~\cite{Kruskal}, which makes the spacetime manifold geodesically complete, does not eliminate this effect of the horizon. However,  Kruskal coordinates tell us what happens behind the horizon, in the region which is inaccessible to us. Analogously, choosing appropriate combinations of the field variables, we can describe the passage through the Big Bang -- Big Crunch singularity, but this does not mean that the presence of such a singularity is not essential. Indeed, extended objects cannot survive this passage.

Furthermore, we have considered early-time and late-time asymptotic solutions for a Bianchi-I universe, filled with a conformally coupled  massless scalar field. As far as we know such solutions were not known before.

To conclude, let us  compare the approach to the problem of the cosmological singularity, developed in this paper and the papers mentioned in the Introduction, with another popular approach to wit attempts to avoid an encounter with the cosmological singularity due to some kind of bounce.  The study of the conditions for  the existence of this kind of bounce has a rather  long history. Let us  briefly mention  some interesting papers devoted to this topic.    In paper~\cite{Parker} it was noticed that in the closed Friedmann model filled with a scalar field the universe can undergo a bounce instead of falling into the Big Crunch singularity. In paper~\cite{Star-bounce} it was shown that the relative number of cosmological evolutions, undergoing the bounce is very small. In paper~\cite{Hawk-bounce} the structure of the set of infinitely bouncing trajectories in such a model was studied, and in paper \cite{Page} it was shown that this set has a fractal nature. The structure of this set of cosmological evolutions and its fractal nature were further studied in papers \cite{we-bounce, we-topol}.  During the last decades, the search for opportunities to avoid  the cosmological singularities is often connected with  quantum cosmology. Quantum effects can modify the effective equations of  cosmological evolution in such a way that  the bounce, which was absent at the classical level, can arise due to these effects. The study of such phenomena is especially extensive in the field of quantum loop cosmology \cite{loop}. In paper \cite{Singh1} it was shown that the semiclassical loop effects can provide the avoidance of the Big Crunch singularity in a closed Friedmann model.   In paper \cite{Singh2} the avoidance of  future singularities in Friedmann models was studied. In paper \cite{Singh3} the bouncing solutions in loop quantum cosmology were analyzed, while in paper \cite{Singh4} a further analysis of the loop quantum cosmology effects in closed Friedmann models was undertaken. The general analysis of cosmological  singularities in loop quantum cosmology was done in paper \cite{Singh5}. The singularities in the Bianchi-I model were studied in \cite{Singh6}. The detailed analysis of the loop quantum cosmology in the Bianchi-I universe was done in paper \cite{MartinBenito}. The paper \cite{Singh7} was devoted to exotic cosmological singularities. In paper \cite{Singh8} some aspects of the loop quantum cosmology of anisotropic Bianchi-I, Bianchi-II and Bianchi-IX models were investigated. A detailed analysis of cosmological singularities in Bianchi-I loop quantum cosmology was done in \cite{Singh9}, while the lack of strong singularities in loop quantum Kantowski-Sachs spacetime was discussed in \cite{Singh10}.
The relation between the Big Bounce and inhomogeneities in loop quantum cosmology was studied in \cite{Brizuela}.
One can say that the loop quantum cosmology produced a general interesting result: In this approach the classical cosmological singularities are replaced by quantum bounces~\cite{Singh01,Singh02,Singh03}.

What is the main difference between the approaches  illustrated  in papers \cite{Parker}--\cite{Singh03} and those, followed in the present paper and in the papers cited in the Introduction? While the authors of the former try to find conditions which allow one  to avoid stumbling upon a cosmological singularity, the authors of the latter attempt to find a way of describing the process of crossing the singularity under consideration. The point is that what looks like a singularity in one set of variables is quite regular in another set.
Furthermore, as shown in papers \cite{Bars1,Bars2,Bars3,Bars4,Bars5,Bars6,Bars7}, \cite{KPTVV-sing} and in the present paper, the crossing of the singularity is accompanied by a switch between
gravity and antigravity regimes. Such a change appears to be  a non-smooth process. Indeed, it is non-smooth in the original variables, but it becomes smooth when considered in terms of the new variables, chosen to describe the singularity crossing.  In all cases such a choice involves some kind of mixing between geometrical and matter variables.
Generally, we think that further investigation of the problems discussed in this paper is of great interest.

\section*{Acknowledgments}
The work of  A.K. was partially supported by the RFBR grant 17-02-01008.
The research of E.P. is supported in part by grant MK-7835.2016.2  of the President of Russian Federation.
The research of S.V. is supported in part by grant NSh-7989.2016.2 of the President of Russian Federation.

\end{document}